\documentstyle[12pt]{article}
\include{psfig}
\newlength{\abstractwidth}
\def\halft{{\textstyle {{{1}\over{2\pi\alpha'}}} }}
\def\quaft{{\textstyle {{{1}\over{4\pi\alpha'}}} }}
\def\half{{\textstyle{1\over2}}}

\begin{document}

\pagestyle{empty}
\begin{titlepage}

\bigskip
\hskip 3.7in\vbox{\baselineskip12pt
\hbox{PSU-TH-206}\hbox{}}
\bigskip\bigskip\bigskip\bigskip

\centerline{\large \bf Path Integral Evaluation of Dbrane Amplitudes} 
\bigskip\bigskip
\bigskip\bigskip

\centerline{\bf 
Shyamoli Chaudhuri\thanks{shyamoli@phys.psu.edu}} 
\medskip
\centerline{Physics Department}
\centerline{Penn State University}
\centerline{University Park, PA 16802}

\bigskip\bigskip

\begin{abstract}
\noindent We extend Polchinski's evaluation of the measure for the one-loop 
closed string path integral to open string tree amplitudes with boundaries 
and crosscaps embedded in Dbranes. We explain how the nonabelian limit of 
near-coincident Dbranes emerges in the path integral formalism. We give a 
careful path integral derivation of the cylinder amplitude including the 
modulus dependence of the volume of the conformal Killing group. 
\end{abstract}

\end{titlepage}

\pagestyle{plain}

\vskip 0.1in
In this brief note we clarify some important issues stemming from the moduli 
dependence of the measure in the open string path integral. We will be primarily 
interested in open string tree amplitudes with multiple boundaries and crosscaps 
on Dbranes. It will suffice for the discussion in this paper to consider the 
simplest configuration of $N$$+$$1$, spatially separated, parallel and static, 
Dpbranes, arrayed in a spatial direction, $X^{25}$, transverse to the branevolume. 
Upon setting the branes in relative motion in an orthogonal spatial direction, 
$X^{24}$, also transverse to the branevolume, the spacetime infrared limit of the
amplitude has an interpretion as the exchange interaction of $N$$+$$1$ Dpbranes
\cite{polchinskiB}. In the supersymmetric theory, its low energy effective 
description is given by supergravity. 

\vskip 0.1in
The spacetime ultraviolet limit of Dbrane amplitudes remains a puzzle in many respects. 
The low energy effective theory in this limit is known to be a $U(1)^{N+1}$ abelian 
gauge theory, with Goldstone bosons arising from the breaking of $X^{25}$ translational 
invariance. These appear at the first massive level in the open string spectrum. In the 
limit of $N$ near-coincident Dbranes, a non-abelian structure is found to emerge 
\cite{polchinski1,polchinskiwitten,witten}, and the massless open string multiplet is 
enhanced from $N+1$ singlets, each transforming as a Lorentz vector on the $p$$+$$1$ 
dimensional Minkowskian branevolume, to a singlet plus the $N^2$-dimensional adjoint 
multiplet of $U(N)$. The additional massless gauge bosons arise from an enhanced background 
of zero length Dirichlet open strings stretched between pairs of {\em distinct} near-coincident 
Dbranes. The open strings have definite orientation, giving a total of $N(N-1)$ distinct 
states. The nonabelian states play an important role in studies of weak/strong coupling 
duality in String Theory. It would be helpful to have a description of their dynamics 
{\em directly} in the world-sheet formalism, beyond that given by the low energy effective 
theory. In what 
follows, we will explain briefly how the nonabelian limit of near-coincident Dbranes 
emerges in the world-sheet formalism. We will also clarify issues relating to the moduli
dependence of the measure in the path integral. A more detailed discussion of the physics
of the ultraviolet limit of open string theory is reserved for future work \cite{asymp}. 

\vskip 0.1in
We begin by clarifying the measure in the open string path integral. The result 
follows rather simply from Polchinski's evaluation of the one-loop closed string 
path integral, a sum over surfaces with the topology of a torus 
\cite{polchinski3}. For surfaces with negative Euler character, with no 
conformal Killing vectors, it is convenient to use uniformization theory 
\cite{bers} to choose a constant curvature gauge slice in the space of 
world-sheet metrics, and a parameterization of the moduli space of metrics that 
leaves the measure for moduli in its simplest form \cite{dhokerphong1}. This procedure 
will be explained in detail for open string tree 
amplitudes in an accompanying paper \cite{asymp}. For surfaces with the topology 
of a cylinder, the conformal Killing volume is a matter of concern. We follow 
the procedure adopted in \cite{polchinski3} and retain the moduli dependence in 
the fiducial world-sheet metric, as opposed to the world-sheet coordinate 
intervals. A careful derivation of the moduli dependence of the conformal 
Killing volume then follows for surfaces of cylindrical topology. This 
procedure correctly recovers both the annulus amplitude with Neumann 
boundaries, and the exchange amplitude for a pair of static Dbranes. The 
extension to the supersymmetric exchange amplitude will be reported upon 
elsewhere \cite{super}.

\vskip 0.1in
Let us begin by expressing the exchange amplitude between $N$ static bosonic 
Dbranes as a conformally invariant path integral over world--sheets. Surfaces 
${\cal M}^{(b,c)}_{h}$ of type $(b,c)$ with $b$ boundaries, $c$ crosscaps, 
and $h$ handles, have Euler characteristic $\chi = 2-2h-(b+c)$, and are 
weighted by a factor $g_{open}^{2\chi}$ in open string perturbation theory. 
To keep things simple, we will for the most part focus on the sum
over orientable world sheets with boundaries, but no handles and/or crosscaps.
In this case, the number of Dbranes, $N$$=$$b+c$, equals the number of 
boundaries, $b$. We follow the method of 
\cite{polyakov} and \cite{polchinski3}, extending to 
string world--sheets with boundaries embedded in Dbranes. The 
separation of parallel Dp-branes in a single spatial 
dimension orthogonal to their worldvolume breaks the $SO(1,d)$ Lorentz 
invariance of the Minkowskian spacetime background to 
$SO(1,p)$$\times$$SO(0,d-p)$. Let ${\bf y}_I$ be a $(d-p)$ 
dimensional vector giving the location
of the $I$th Dbrane. We have the boundary conditions:
\begin{eqnarray}
\label{bndrycondi}
X^m_{(I)} &=& y^m_I, \quad I= 1, \cdots , b ; ~ m=p+1, \cdots , d \cr
n^a_{(I)} \partial_a X^{\mu}_{(I)} &=& 0, \quad \mu=0, \cdots p \quad ,
\end{eqnarray}   
where ${\hat n}_I$ is an inward pointing normal to the $I$th boundary
circle, $C_I$, $I$$=$$1, \cdots , b$. The relative locations of the 
Dbranes are assumed fixed. In the T--dual picture, 
the ${\bf y}_I$ correspond to Wilson lines wrapped around
the compact dual coordinates, ${\tilde {\bf X}}$, of the critical bosonic 
open string theory compactified on a $d-p$ dimensional torus 
\cite{polchinskiB}. 

\vskip 0.1in
The exchange amplitude can be expressed as a Polyakov path integral 
\cite{polyakov} over orientable Riemann surfaces, ${\cal M}$, with 
boundaries, $C_I$, $I= 1, \cdots , b$, embedded in parallel Dpbranes:
\begin{equation}
{\cal A} = \int
{{[d\delta X][d\delta g]}\over{Vol[{\rm Diff}\times{\rm Weyl}]}}
             e^{ - \left [ 
   \quaft \int_{\cal M} d^2 \xi  {\sqrt {g}} { g}^{ab}  
        \partial_a {X} \partial_b {X} ~+~ 
 S_{\rm ren.}[ g; \nu_0,\mu_0, \lambda_0^{(I)}] \right ] }  \quad ,
\label{pathi}
\end{equation}
on which we must impose the boundary conditions in eq.(1). The bare action 
includes all local, renormalizable, terms necessary to ensure exact conformal 
invariance of the path integral. Thus, $S_{\rm ren.}$
includes bulk, and boundary, cosmological constants, plus a term proportional 
to the Euler characteristic of the surface:
\begin{equation}
S_{\rm ren.} =  {{\nu_0}\over{2 \pi}} [ \half \int_{\cal M} d^2 \xi 
        {\sqrt{ g}} R_{ g} + \sum_{I=1}^b \oint_{C_I} ds 
            { \kappa}  ] +  \mu_0 \int_{\cal M} d^2 \xi {\sqrt{ g}}  + 
          \sum_{I=1}^b \lambda_0^{(I)} \oint_{C_I} ds \quad .
\label{ren}
\end{equation}
The renormalization constants, $[\nu_0,\mu_0;\lambda^{(I)}_0]$,  will be 
cancelled in the critical spacetime dimension by divergent counterterms 
originating in the Weyl anomaly of the measure. The factor in square 
brackets in eq.(3) is a topological invariant equal to the Euler 
characteristic, $2 \pi \chi$, from the Gauss--Bonnet theorem, and
${\hat \kappa}$ is the geodesic curvature on the boundary. The gauge 
fixed path integral in eq.(2) is required to be exactly invariant under 
conformal reparameterizations of the surface including boundaries. 
We will use this as a guiding principle in determining the measure
for the path integral.

\vskip 0.1in
Let $C_I$ denote the $I$th boundary circle with parameterization unspecified;
we refer to this as a {\em physical} boundary circle. Recall that an 
arbitrary boundary metric, also known as an einbein, $e(\lambda)$, 
can be brought to a constant by a reparameterization, 
$\lambda \to f(\lambda)$, where $\lambda$ is the circle 
variable, $0\le \lambda \le 1$, and the only coordinate invariant property 
of the circle, $C_I$, is its physical length, $l_I$$=$$ \int_0^1 d \lambda 
{e}(\lambda) $. Let ${\hat g}_{ab}(\xi;\tau)$ denote a world sheet metric with 
fixed conformal class characterized by moduli, $\tau$, with corresponding 
einbein, ${\hat e}_I(\lambda;l_I)$$=$${\sqrt {\hat g}}(\xi;\tau)|_{C_I}$, 
on the $I$th boundary circle. Physics requires us to sum over all world 
sheets linking the physical boundary circles in the path integral. We will 
make this gauge fixing procedure on world sheet metrics explicit below.

\vskip 0.1in
We begin with the integration over coordinate embeddings, $X^M(\xi)$, 
with a fixed fiducial metric, ${\hat g}_{ab}(\xi;\tau_i)$, on the world
sheet, and fixed einbeins on the boundary circles, 
${\hat e}_I$$=$$ {\sqrt {\hat g}}|_{C_I}$, for all $I$$=$$1, \cdots , b$. 
The metric in the tangent space for small variations, 
$(\delta X^m,\delta X^{\mu})$, $m$$=$$p+1, \cdots , d$, $\mu=0, \cdots , p$,
is required to be reparameterization invariant. Normalizing the Gaussian 
integral over infinitesimal variations to unity, we can factor out and 
perform the integration over constant modes $\delta X_0^{\mu}=\delta X^{\mu} 
- \delta X^{'\mu}$, $\mu=0, \cdots p$, where the primes denote nonconstant 
modes \cite{polchinski3}. Since we fix the location of the branes in the 
Dirichlet directions, there are no constant modes to be integrated over. We 
regulate the infrared divergence coming from the integration over $p+1$ 
noncompact directions parallel to the branevolume 
by putting the system in a box of volume $V_{p+1}$: 
\begin{equation}
\prod_{\mu=0}^p \int [d \delta {\bf X}_0 ] = V_{p+1} \quad .
\label{zeromode}
\end{equation}

\vskip 0.1in
For the Dirichlet directions, we change basis to the $b-1$ distances,
$|{\bf y}_I-{\bf y}_J|$, $I$$\neq$$J$, and the center of mass, 
${\bf y}_{\rm c.m.}$, each a $d-p$ dimensional vector. For spatially 
separated branes, we must include the contribution to the classical 
action, $T_{(IJ)}$, for every $I$$\neq$$J$, from open strings stretched 
between the $(IJ)$th pair of branes. Thus, the static exchange amplitude only 
depends upon the distances between branes. We can integrate over the center 
of mass of the brane configuration to obtain the volume of the orthogonal 
$(d-p)$ dimensional Dirichlet space, $V_{d-p}$.

\vskip 0.1in
Performing the Gaussian integrations over zero modes in the Neumann
directions gives the 
normalization of the measure for nonconstant modes:
\begin{equation}
\label{measureex}
\prod_{\mu=0}^p \int [d \delta X^{'\mu} ] e^{- \quaft |\delta X'|^2} = 
       [(4\pi^2\alpha')^{-1}\int d^2 \xi {\sqrt{\hat g}}]^{(p+1)/2} 
\quad ,
\end{equation}
with the analogous integration in the Dirichlet directions normalized to
unity \cite{polchinski3}.
We can expand the infinitesimal nonconstant variations, $\delta X^{'M}$, 
in a complete set of harmonic functions satisfying the appropriate 
boundary condition on the $C_I$. Then the Gaussian integrations over 
coordinate embeddings can be straightforwardly performed with the result:
\begin{equation}
\label{funcemb}
V_{d+1} (4\pi^2\alpha')^{-(p+1)/2} e^{- \sum_{I\neq J} T_{IJ}} 
   (\int d^2 \xi {\sqrt {\hat g}})^{(p+1)/2} 
        ({\rm Det}' \Delta_0 )^{-(d-p)/2} ({\rm Ndet}' \Delta_0 )^{-(p+1)/2}
\quad ,
\end{equation}
where Det and Ndet denote the functional determinants evaluated, respectively,
with Dirichlet and Neumann boundary condition on the $C_I$. On the cylinder 
these functional determinants are identical.

\vskip 0.1in
The integration over world--sheet metrics is treated as in Polchinski's
evaluation of the measure for the one--loop path integral on the torus
\cite{polchinski3}, except that we specialize to a fiducial metric with
constant curvature. The number of moduli, $n_m$, and 
conformal Killing vectors, $n_c$, are as given by the Riemann--Roch theorem:
$3\chi$$=$$n_c$$-$$n_m$. A specific choice of basis for 
quadratic differentials on the Riemann surface 
corresponds to a specific parameterization, $[\tau_i]$, of conformally 
inequivalent metrics ${\hat g} (\xi;\tau_i)$. The real 
parameters $\tau_i$, $i=1, \cdots n_{\rm m}$, are the moduli of the 
Riemann surface, $b$ of which can be related to the lengths, $l_I$, 
$I= 1, \cdots , b$ of the boundaries as measured by the fiducial 
boundary metric, ${\sqrt {\hat g}}(\xi;\tau_i)|_{C_I}={\hat e}_I(\lambda;l_I)$. 
The remaining $2(b-3)$ moduli have a simple interpretation in terms of 
lengths and angles parameterizing the shape of internal geodesics on the 
constant curvature surface. The nonorientable surface, ${\cal M}_{(b,c)}$,
can be obtained from the orientable surface, ${\cal M}_{(b+c,0)}$, with
$b+c$ boundaries, by ``plugging'' $c$ holes with crosscaps. The counting 
of moduli is straightforward in this gluing. Since the boundary of the 
crosscap is identified with the boundary of the hole upto a relative
twist, we lose the free length parameter for the geodesic boundary of the
hole, gaining an angle from the relative twist of crosscap to hole. 
Thus, for surfaces of negative Euler character 
with both boundaries and crosscaps, $n_m$$=$$b+c$$+$$2(b+c-3)$. 
Making the appropriate extensions to the analysis of \cite{polchinski3} we 
obtain the result:
\begin{eqnarray}
\label{meassy}
\int {{[d \delta g]}\over{Vol[{\rm Diff}\times{\rm Weyl}]}}  &\to&  
 \quad  (2\pi)^{(n_{\rm c}-n_{\rm m})/2}  
\prod_{i=1}^{n_m} \int d \tau_i ({\rm Det}'\Delta_1)^{1/2} 
 (\int d^2 \xi {\sqrt {\hat g}})^{{\rm n_m}/2} \cr 
        \quad & & ( {\rm det}Q_{ab})^{-1/2} 
       [{\rm det } ((\zeta_k)_{ef} (\zeta_l)^{ef})]^{1/2}     
                 \quad .
\end{eqnarray}
The notation for the various factors in the Jacobian is as follows. 
$\Delta_1$ is the
Laplacian acting on vector fields on the Riemann surface, related to 
the scalar Laplacian by the identity: $(\Delta_1)_c^d$$=$$-\delta_c^d 
\Delta_0$$- $$\nabla^d \nabla_c $$+$$\nabla_c\nabla^d$. 
$({\rm det} Q_{ab})^{-1/2}$
is the contribution to the Jacobian from the constant modes of the vector
Laplacian. The matrices, $\zeta_i$, $i$$=$$1, \cdots , n_m$ are determined 
by the moduli dependence of the fiducial metric, 
$(\zeta_i)_{ab}$$=$${\hat g}_{ab,i}$$-$$\half {\hat g}_{ab} {\hat g}^{cd}
{\hat g}_{cd,i}$. Thus, given an explicit form for the constant curvature
metric with dependence on $n_m$ real moduli parameters manifest, the 
measure for the path integral is completely determined. In the critical 
spacetime
dimension, the integration over the volume of the group of reparameterizations
continuously connected to the identity, ${\rm Diff}_0$$\times$${\rm Weyl}$,
can be straightforwardly performed leaving the result on the L.H.S. of
eq.(8).
For surfaces with $(b+c)$$\ge$$3$, the contribution from the conformal 
Killing vectors is to be dropped from this expression since $n_c$$=$$0$. 

\vskip 0.1in
Thus, the expression for the gauge fixed path integral takes the form:
\begin{equation}
\label{pathd}
{\cal A} = \int [d \tau] e^{- \sum_{I\neq J} T_{IJ}} 
     ( {\rm Det}' \Delta_1)^{1/2} ({\rm Det}' \Delta_0 )^{-(d-p)/2}
        ({\rm Ndet}' \Delta_0 )^{-(p+1)/2} \quad ,
\end{equation}         
where the normalized measure for moduli, $[d \tau]$, is given by:
\begin{eqnarray}
\label{measmod}
[d \tau] &=&   V_{d+1} (2\pi)^{(n_{\rm c}-n_{\rm m})/2}
          (4\pi^2\alpha')^{-(p+1)/2}  
      \prod_{i=1}^{n_m} d \tau_i 
 (\int d^2 \xi {\sqrt {\hat g}})^{({\rm n_m}+(p+1))/2} \cr
\quad & &  ( {\rm det}Q_{ab})^{-1/2}
       [{\rm det } ((\zeta_k)_{ef} (\zeta_l)^{ef})]^{1/2}     
                 \quad .
\end{eqnarray}

\vskip 0.1in 
To proceed further, we need an explicit parameterization of the Riemann 
surface following global uniformization to some region in the complex 
plane. An explicit parameterization of this region, and the formulation 
of the eigenfunction problem, is a challenging problem for Riemann surfaces 
with $b+c\ge3$ boundaries and/or crosscaps. We reserve that discussion for 
future work \cite{asymp}. The disk is a special case all by itself because
it uniformizes 
to the unit circle with positive constant curvature metric; it also has 
three conformal Killing vectors. 
Surfaces with cylindrical topology are of course easiest in this respect 
since they can be conformally mapped to a rectangle in the flat complex 
plane, and the eigenfunction 
problems with either choice of boundary condition have an explicit solution. 
The cylinder also has a conformal Killing vector so it is helpful to treat
it separately, as we do now.  We will recover both the 
annulus diagram of open string theory with Neumann boundaries and the 
exchange amplitude between 
a pair of static Dbranes \cite{polchinskiB}.

\vskip 0.1in
An arbitrary cylinder can be uniformized to a 
rectangle in the complex plane with area: 
\begin{equation}
{\rm A} = \int d^2\sigma {\sqrt{\hat g}} =t \quad .
\label{bulk}
\end{equation}
We choose a parameterization such that the rectangle is bounded by the 
unit intervals, $0\le \sigma^1 \le 1$, $ 0 \le \sigma^2 \le 1$. Then
the moduli dependence is restricted to the flat world--sheet metric:
\begin{equation}
d s^2 = (d \sigma^1)^2 + t^2 (d \sigma^2)^2 \quad ,
\label{cylmets}
\end{equation} 
and $t$ also corresponds to the length of the cylinder. The complete 
set of eigenfunctions of the scalar Laplacian on this domain are simply 
the circular functions of two variables, $(\sigma^1, \sigma^2)$. 
The length of the cylinder, $t$, is an Euler--Lagrange parameter
appearing in the effective action. Note that, for the special case 
of the cylinder, renormalizations of the bulk and boundary cosmological 
constants are not independent. The cylinder has a single conformal Killing 
vector, $\eta_0$, which 
contributes to the path integral a $1 \times 1$ matrix with determinant, 
$(2\pi/{\rm det} Q_{ab})^{1/2}$$=$$(2\pi/t)^{1/2}$. 
The measure for moduli can be computed from the matrix, 
$(\zeta)_{ab}$$=$${\hat g}_{ab,t} - \half {\hat g}_{ab} {\hat g}^{cd}
      {\hat g}_{cd,t} $. It takes diagonal form on the cylinder, 
$\zeta_{11}=-1/t$, $\zeta_{22}=t$, and has determinant, 
$({\rm det} \zeta^{ab} \zeta_{ab})^{1/2}$$=$$1/t$. 

\vskip 0.1in
By a reparameterization of the worldsheet, the unit normal and unit 
tangent vectors at both boundary circles, $C_1$, $C_2$, 
can be chosen to lie along the $(\sigma^2,\sigma^1)$ grid.
The Neumann determinant is composed 
from the basis of periodic functions nonvanishing on the boundary, 
$\sigma^2$$=$$0$, $1$:
\begin{equation}
\psi^{(1)}_{(n_1,n_2)}(\sigma^1, \sigma^2) = 
 e^{2 n_1 \pi i \sigma^1} {\rm Cos} ({ n_2 \pi \sigma^2}) \quad , 
\label{modeneu}
\end{equation}
with $-\infty $$\le $$ n_1 $$ \le $$ \infty$, and $n_2 > 0$. 
The Dirichlet determinant is composed from the orthogonal basis of 
periodic functions vanishing on the boundary:
\begin{equation}
\psi^{(2)}_{(n_1,n_2)}(\sigma^1, \sigma^2) = 
 e^{2 n_1 \pi i \sigma^1} {\rm Sin} ({ n_2 \pi \sigma^2}) \quad , 
\label{modedir}
\end{equation}
once again, with $-\infty $$\le $$ n_1 $$ \le $$ \infty$, and $n_2 > 0$. 
Either choice of basis satisfies a completeness relation on the 
cylinder.

\vskip 0.1in
Substituting $n_c$$=$$n_m$$=$$1$ in eq.(10) for the measure, $[d \tau]$,
we recover the familiar result:
\begin{equation}
\int [d \tau] = \int {{dt}\over{t}} ~ V_{d+1} (4\pi^2 \alpha')^{-(p+1)/2}
              t^{(p+1)/2} \quad , 
\label{modcyl}
\end{equation}
and the gauge fixed path integral takes the form: 
\begin{equation}
\label{cylamp}
{\cal A}(y) = \int {{dt}\over{t}}  V_{d+1} (4\pi^2\alpha')^{-(p+1)/2} 
      t^{(p+1)/2} e^{- \quaft y^2/t }
             [ {\rm Det}' \Delta_0]^{-12} \quad ,
\end{equation}
where we have substituted $d-1$$=$$24$ for the critical string, and the 
relation between vector and scalar Laplacians, 
$({\rm Det}' \Delta_1)^{1/2}$$=$${\rm Det '}\Delta_0$, which holds on the 
cylinder. The classical contribution to the action is from open strings 
stretched between the branes.

\vskip 0.1in
The functional determinants on the cylinder can be evaluated directly 
following the analysis in the appendix of ref.\cite{polchinski3}. The 
eigenmodes of the scalar Laplacian with zero Dirichlet 
boundary condition, $\delta \eta(\sigma^1)|_{\sigma^2=0,1}=0$, are composed
from the basis functions in eq.(14), and the unregulated determinant 
is therefore the infinite product:
\begin{equation}
{\rm Det}' \Delta_0 = \prod_{n_1,n_2}
 \left [ {{4 \pi^2}\over{t^2}} (n_1^2 t^2 + {{n_2^2}\over{4}} ) \right ] 
\quad , 
\label{unregdet}
\end{equation}
with the restrictions, $-\infty \le n_1 \le \infty$, $n_2>0$. Equivalently, 
we could compute the product over eigenmodes of the vector Laplacian with 
zero Dirichlet condition, i.e., the variations $\eta^a(\sigma^1)|_{\sigma^2
=0,1} = 0$. This gives the unrestricted product, 
$- \infty $$ \le $$ n_1 $$ \le $$ \infty$, 
$- \infty $$ \le $$ n_2 $$ \le $$ \infty$, with the $n_1=n_2=0$ term 
subtracted out. The 
required scalar determinant is obtained by taking its square root 
\cite{polchinski3}\cite{cohen}. The result for the determinant of the 
scalar Laplacian with Dirichlet condition on the cylinder is \cite{cohen}:
\begin{equation}
({\rm Det}'  \Delta_0 )^{1/2} = (2t)^{\half} e^{-2\pi t/12} \prod_{n=1}^{\infty}
 (1 - e^{-4 n \pi t}) = (2t)^{\half} \eta(2it) = \eta(i/2t) \quad .  
\label{determs}
\end{equation}
where we have used a modular transformation in the variable $t$ to 
obtain the second equality. From eqs.(13) we have an identical answer 
for the Neumann determinant.
Substituting in eq.(16), the static amplitude between parallel Dpbranes 
is given by the expression:
\begin{equation}
{\cal A}(y) = \int_0^{\infty} {{dt}\over{t}} V_{d+1} 
   (4\pi^2\alpha')^{-(p+1)/2} t^{(p+1)/2} e^{-\quaft y^2/t }  
      [\eta(i/2t)]^{-24} \quad .
\label{ampd}
\end{equation}
A change of variables, $s$$=$$1/2t$, in the integral gives the equivalent
form:
\begin{equation}
{\cal A}(y) = \int_0^{\infty} {{ds}\over{s}} V_{d+1} 
   (8\pi^2\alpha's)^{-(p+1)/2} e^{-\halft y^2s } [\eta(is)]^{-24} \quad .
\label{ampdtwo}
\end{equation}
which can be compared with the expression for the exchange amplitude 
between static bosonic Dpbranes obtained in the operator formalism 
\cite{polchinskiB}. The variable $s$ corresponds to the length of
the boundary. Setting $p$$=$$d$ we recover the cylinder amplitude 
of open string theory with Neumann boundaries, in a single Chan--Paton
state, as originally obtained in \cite{frsuss}. This is also consistent 
with the result from the method of images \cite{nielsen}, which can be 
used to relate the cylinder and torus amplitudes. 

\vskip 0.1in
We now return to the puzzle mentioned in the introduction-- the emergence of 
nonabelian structure in the world-sheet picture, in the limit of 
near-coincident Dbranes. A simple nonabelian configuration with an interesting 
low energy 
description is that of a single probe-Dbrane, spatially distant from a pair 
of near-coincident Dbranes. It can be given a world-sheet description as 
follows. Consider a {\em pants} surface: an orientable Riemann surface with 
boundaries on three Dbranes, with boundaries $C_2$, $C_3$ on near-coincident 
Dbranes. The low-energy effective theory on the branevolume is a 
$[SU(2)$$\times$$U(1)]$$\times$$U(1)$ nonabelian gauge theory. 

\vskip 0.1in
A generic pants surface can be mapped to a simply connected domain as follows. 
Pick a base point on the surface, $z_0$, and cut along paths joining $z_0$ 
to the three boundaries. The resulting surface can be mapped to a $-9\chi$-sided
polygon, a simply connected domain in the complex plane. For a constant 
curvature pants surface, with ${\hat R}$$=$$-1$, this gives a polygon in 
the upper half plane with hyperbolic metric \cite{bers}, and the joining 
paths can be chosen as geodesics on the surface. Going around the polygon 
once, we can label the edges: 
\begin{equation}
\label{pantsdom}
(r_1,s_1,{\tilde s}_2,r_2,s_2,{\tilde s}_3,r_3,s_3,{\tilde s}_1) \quad ,
\end{equation}
where the $r_i$ are the images of the $C_i$ upon mapping to the complex plane.
The images of the base point lie at the intersections of adjacent pairs of
joining curves, ($s_i$,${\tilde s}_{i+1}$), $i$$=$$1$, $\cdots$, $3$, taken
in cyclic order. Notice that the location of the base point {\em within}
the surface is arbitrary, but the limit of the mapping when the base point 
approaches any boundary is {\em singular}: the domain contains a cusp on its
boundary, and the ordinary properties of function theory on the domain have 
to be suitably modified to take into account the cusp. 

\vskip 0.1in
What is the shape of the world--sheets that contribute to the singular limit
of this mapping? Consider the case that $z_0$ lies on boundary $C_3$, and 
the pair, ($s_3$,${\tilde s}_3$), is shrunk to zero length. Now the boundary 
lengths, $l_i$, correspond to open string proper times. Recall that the 
inverse length, $l_{[12]}$$=$$1/l$, of a cylinder can be interpreted, through 
open-closed string world-sheet duality, as the proper time of a propagating 
closed string exchanged between a pair of Dbranes \cite{polchinskiB}. A 
singular pants surface can be given a surprisingly simple interpretation in 
the language of closed string "proper times". The singular world--sheet can 
be visualized as a closed string emitted by Dbrane $\bf D_1$, which propagates 
smoothly towards Dbrane $\bf D_2$. In addition, $\bf D_3$ emits a closed string
which propagates for a vanishingly short proper time, $l_{[23]}$, before being 
absorbed by Dbrane $\bf D_2$. Consider the limit when boundary $C_2$ approaches
boundary $C_3$, as with near-coincident branes. Such a pants surface can be 
mapped to the complex plane by cutting along only {\em two} joining curves, 
with the base point $z_0$ on the boundary $C_3$. The apparent mismatch in the
counting of moduli is accounted for by the presence of the cusp, in agreement
with the Gauss-Bonnet and Riemann-Roch theorems.  

\vskip 0.1in 
Function theory on the resulting domain has to be modified by extending the 
analysis of the eigenfunction problem to domains with a cusp, a particular 
instance of a domain with an isolated singular point \cite{couranthilbert2}. 
We will reserve that discussion for future work \cite{asymp}.

\vskip 0.2in
\noindent{\large {\bf Acknowledgments}}

\vskip 0.1in
I would like to thank my spouse for enabling me to finish this work.
I also thank my students, Y. Chen and E. Novak, for helpful comments 
on the draft. 

\vskip 0.1in
\noindent{\bf Note added:} The understanding of singular pants surfaces given
here was worked out in late November 1998. I would like to thank the referee 
for the suggestion that I include a discussion of the nonabelian limit of 
near-coincident Dbranes.

\end{document}